\documentclass{iau}
\usepackage{graphicx}
\usepackage{natbib}
\newcommand{\kms}{$\mathrm{km\, s^{-1}\, }$}

\newcommand{\mbh}{M_{\bullet}}

\def\apj{ApJ}%
\def\apjl{ApJ}%
\def\aap{A\&A}%
\def\mnras{MNRAS}%
\def\pasp{PASP}%
\def\nat{Nature}%
\title[IMBHs in globular clusters] 
{Intermediate-mass black holes in globular clusters: observations and simulations}

\author[Nora L\"utzgendorf et al.]   
{Nora L\"utzgendorf$^1$, 
 Markus Kissler-Patig$^2$, 
 Karl Gebhardt$^3$, 
 Holger Baumgardt$^4$,
 Diederik Kruijssen$^5$,
 Eva Noyola$^3$,
 Nadine Neumayer$^6$,
 Tim de Zeeuw$^{7,8}$,
 Anja Feldmeier $^7$,
 Edwin van der Helm$^8$,
 Inti Pelupessy$^8$,
 \and Simon Portegies Zwart$^8$}

\affiliation{$^1$ESA, Space Science Department,\\
              	Keplerlaan 1, NL-2200 AG Noordwijk, The Netherlands\\ 
              	email: {\tt nluetzge@cosmos.esa.int} \\[\affilskip]
              
				$^2$Gemini Observatory, Northern Operations Center, \\
             	670 N. A'ohoku Place, Hilo, Hawaii, 96720, USA\\
				
				$^3$Department of Astronomy, University of Texas at Austin, \\
			    Austin, TX 78712, USA \\
			    
			    $^4$School of Mathematics and Physics,\\
			 	University of Queensland, Brisbane, QLD 4072, Australia\\
			 	
				$^5$Max-Planck Institut f\"ur Astrophysik,\\
                Karl-Schwarzschild-Stra\ss e 1, D-85748, Garching, Germany \\ 			 	
			 	
			 	$^6$Max-Planck-Institute for Astronomy, \\
			 	K\"onigstuhl 17, 69117, Heidelberg, Germany	 \\
			 	
				$^7$European Southern Observatory,\\
              	Karl-Schwarzschild-Stra\ss e 2, D-85748 Garching, Germany\\	 	
			 	
			 	$^8$Leiden Observatory, Leiden University,\\
			 	PO Box 9513, NL-2300 RA, Leiden, The Netherlands
				}

\pubyear{2015}
\volume{312}  
\pagerange{119--126}
\jname{Star clusters and black holes in galaxies across cosmic time}
\editors{A.C. Editor, B.D. Editor \& C.E. Editor, eds.}
\begin{document}

\maketitle

\begin{abstract}
The study of intermediate-mass black holes (IMBHs) is a young and promising field of research. If IMBH exist, they could explain the rapid growth of supermassive black holes by acting as seeds in the early stage of galaxy formation. Formed by runaway collisions of massive stars in young and dense stellar clusters, intermediate-mass black holes could still be present in the centers of globular clusters, today. Our group investigated the presence of intermediate-mass black holes for a sample of 10 Galactic globular clusters. We measured the inner kinematic profiles with integral-field spectroscopy and determined masses or upper limits of central black holes in each cluster. In combination with literature data we further studied the positions of our results on known black-hole scaling relations (such as $\mbh - \sigma$) and found a similar but flatter correlation for IMBHs. Applying cluster evolution codes, the change in the slope could be explained with the stellar mass loss occurring in clusters in a tidal field over its life time. Furthermore, we present results from several numerical simulations on the topic of IMBHs and integral field units (IFUs). We ran N-body simulations of globular clusters containing IMBHs in a tidal field and studied their effects on mass-loss rates and remnant fractions and showed that an IMBH in the center prevents core collapse and ejects massive objects more rapidly. These simulations were further used to simulate IFU data cubes. For the specific case of NGC 6388 we simulated two different IFU techniques and found that velocity dispersion measurements from individual velocities are strongly biased towards lower values due to blends of neighbouring stars and background light. In addition, we use the Astrophysical Multipurpose Software Environment (AMUSE) to combine gravitational physics, stellar evolution and hydrodynamics to simulate the accretion of stellar winds onto a black hole. 
\keywords{stars: kinematics and dynamics, methods: numerical, black hole physics}
\end{abstract}

\firstsection 
\section{Introduction}

Intermediate-mass black holes (IMBHs) provide the missing link between supermassive black holes and stellar-mass black holes. Their masses range from a few hundred solar masses up to $10^5 M_{\odot}$. Possible formation scenarios are remnants of population III stars \citep{madau_2001} or runaway merging in young dense star clusters \citep[e.g.][]{zwart_2004}. For supermassive black holes, correlations between the black-hole mass ($M_{\bullet}$) and several properties of their host system, such as velocity dispersion \citep[$\sigma$, e.g. ][]{ferrarese_2000} or total mass \citep[$M$, e.g. ][]{haering_2004} are observed. Extrapolating these correlations to low mass systems ($\sigma \sim 10 - 30$ \kms) suggests that IMBHs could reside in today's massive globular clusters. Detecting and measuring these black holes would provide important data points to the $M_{\bullet}-\sigma$ relation at the lower mass end. Furthermore, IMBHs could explain the rapid growth of supermassive black holes at high redshift by acting as seeds in the early universe. 

Many attempts have been made to detect IMBHs in globular clusters. These range from spectroscopic and photometric velocity measurements \cite[e.g.][]{noyola_2008} to X-ray and radio observations in order to find signatures of accretion \citep[e.g.][]{strader_2012}. The results are disputed and the question whether IMBHs exist in globular clusters is not yet resolved. All methods bring their caveats. The low gas content in globular clusters makes the detection of accretion signatures rather difficult. The X-ray and radio signals from low and irregular accretion rates are not yet well understood. On the other hand, kinematic signatures suffer from shot noise caused by a few bright stars or contamination from background light. It is therefore crucial to advance the field of IMBHs in all directions. Simulations on globular clusters and observing techniques provide a valuable tool to verify observing results and to understand the internal processes of globular clusters. 

Previous numerical work on IMBHs in globular clusters has been performed by \cite{baumgardt_2005} and \cite{noyola_2011} who found that the surface brightness profiles of clusters hosting an IMBH exhibit weak central cusps, in contrast to core-collapsed clusters with very steep profiles and pre-core collapsed systems with no cusp at all. However, \citet{trenti_2010} and \citet{vesperini_2010} showed that the shallow cusp may also form as a transition state of globular clusters undergoing core collapse and therefore cannot be a sufficient criterion for a globular clusters hosting an IMBH. Furthermore, it has been shown that IMBHs prevent a cluster from undergoing  core collapse and reduce the degree of mass segregation compared with non-IMBH clusters \citep{baumgardt_2004b,gill_2008}. \cite{trenti_2007} predicted that clusters with a high ratio of core radius to half-mass radius are good candidates for hosting an IMBH. This was challenged by \cite{hurley_2007}, who showed that the ratios observed for Galactic globular clusters can be explained without the need for an IMBH when treating model data as if they were observational data.



\section{Observations}

We used integral field spectroscopy provided by the FLAMES \citep[Fiber Large Array Multi Element Spectrograph,][]{pasquini_2002} instrument mounted on UT2 at the Very Large Telescope (VLT) to obtain integrated light spectra from 10 Galactic globular clusters. These clusters have been selected by their mass and the shape of their density profiles to be good candidates for hosting IMBHs. $N$-body simulations have shown that an IMBH in the core of a globular cluster prevents the cluster to undergo core collapse and produces a large core with a shallow cusp in its density profile \citep{baumgardt_2005}. 

In addition to the spectroscopic data we used HST images from the archive for each cluster. From these images we obtained the color magnitude diagram, the photometric center and the surface brightness profile. The center determination is a crucial step as the shape of the surface brightness profile and the kinematic profile depend on its position. We therefore used several methods to determine the center and to confirm our results \citep[e.g. isodensity contours, pie wedges][]{nora11}. Another important part of the photometry is the surface-brightness or density profile of the globular cluster. This is used as an input for the dynamical modelling. The profile is obtained by applying radial bins around the photometric center and computing the density/brightness in each bin with a combination of star counts and pixel statistics. 

For the spectroscopic data, the large integral field unit ARGUS was pointed at the center of each globular cluster and if necessary a mosaic was produced to cover a substantial area of the core radius. The three-dimensional data cube of the observations was used to construct a velocity map to check for rotation or spurious kinematic features. To compare with dynamical models, however, a velocity dispersion profile is needed. To obtain this, radial bins were applied around the photometric center and the spectra falling in these bins were combined. From the combined spectrum we measured the line broadening by cross correlating with a template spectrum which gives the direct value of the velocity dispersion from the integrated light. The uncertainties for each velocity dispersion measurement were acquired by running Monte Carlo simulations on the simulated IFU using the information from the high-resolution HST image. 
\begin{figure}[h!]
\begin{center}
 \includegraphics[width=3.1in]{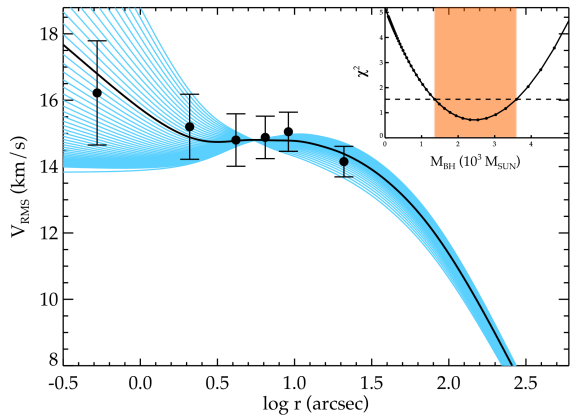} 
 \caption{Velocity-dispersion profile of NGC 6266 overplotted by Jeans models with different black-hole masses. The $\chi^2$ values is shown in the upper right and the best fit model indicated by the black sold line.}
   \label{fig:model}
\end{center}
\end{figure}
 
The resulting velocity dispersion profile was then used to determine the possible existence and mass of an intermediate-mass black hole in the center. For this, the derived density profile of each cluster was used as an input for Jeans models. These models, derived from the collisionless Boltzmann equation, use the density profile of a stellar system to predict the second moment of the velocity distribution i.e., the velocity dispersion. A central mass, such as a central black hole can be added to the model. By applying $\chi^2$ statistics we found the model that fits the data best. Figure \ref{fig:model} shows the fit of the globular cluster NGC 6266 with an IMBH signature. The profile clearly rises and requires a model with a black-hole mass of $\sim 3000 \ M_{\odot}$. This method was applied to all clusters in the sample and for each one we reported an upper limit or a value on the black hole mass \citep{nora11,nora12a,nora13,feldmeier_2013}.

\begin{figure}[h!]
\begin{center}
 \includegraphics[width=5.0in]{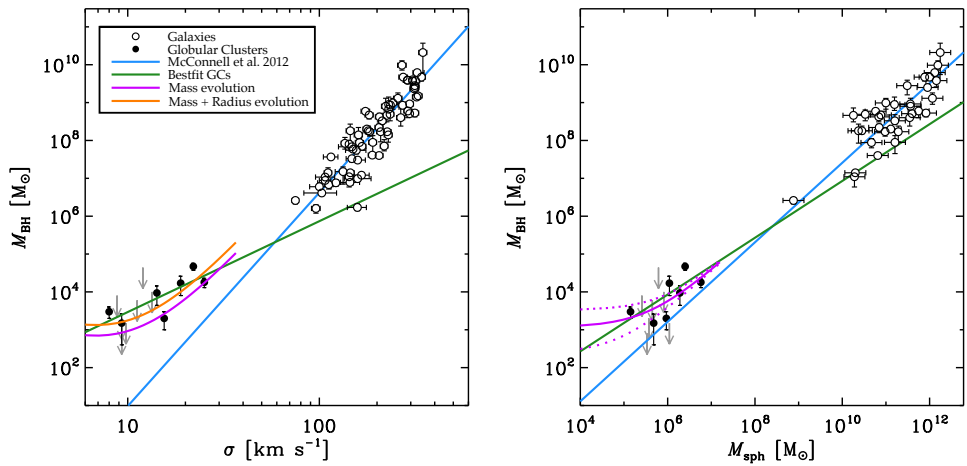} 
 \caption{$\mbh-\sigma$ and $\mbh-M$ relation for supermassive black holes and IMBHs. The clear offset and shallower slope is well reproduced by cluster evolution models considering mass evolution only (magenta line) as well as models that additionally take the radius evolution (i.e. expansion) into account \citep[orange line][]{kruijssen_2013}. }
   \label{fig:msig}
\end{center}
\end{figure}

Using the data points and upper limits of our sample together with measurements from the literature, we compared the correlation between black-hole mass and host system properties such as velocity dispersion and total mass with those measured for supermassive black holes in galaxies. In \cite{nora13b} we analyzed this small and challenging data set by applying survival analysis in combination with Markov Montecarlo Chains in order to account for uncertainties and upper limits. We computed correlation coefficients and linear regressions for the major correlations such as $\mbh-\sigma$, $\mbh-L$ and $\mbh-M$ as well as possible dependencies on half-mass radius, metallicity and galactocentric distance. We found that the major correlations are prominent but shallower than those for supermassive black holes (Figure \ref{fig:msig}). In a follow up paper \citep{kruijssen_2013} we showed that a possible explanation for the offset more shallow correlation is the severe mass loss and expansion of globular clusters during their life time in a tidal field which leads to a reduction of the velocity dispersion and total mass of the system (see Figure \ref{fig:msig}).

\section{Simulations}

Because of large uncertainties in the measurements and contradicting results from different methods in the field of IMBHs, it is crucial to support the observing strategies, analysis and physics of the observations with sophisticated simulations. The next sections describe the different simulations that were performed to compare observations and verify our observing techniques. 

\subsection{N-body simulations}

$N$-body simulations are a valuable tool to understand the physics of self gravitating systems such as globular clusters and to identify possible observables of IMBHs in their centers. In \cite{nora13c} we ran $N$-body simulations based on the GPU (Graphic Processing Unit)-enabled version of the collisional $N$-body code NBODY6 \citep{aarseth_1999,nitadori_2012} on GPU graphic cards at the Headquarters of the European Southern Observatory (ESO) in Garching and the University of Queensland in Brisbane. This code uses a Hermite integration scheme with variable time steps. Furthermore, it treats close encounters between stars by applying KS \citep{KS} and chain regularizations and accounts for stellar evolution \citep{hurley_2000}. The regularization procedures are crucial for following orbits of tightly bound binaries over a cluster lifetime accurately and treat strong binary-single and binary-binary interactions properly. The simulations were carried out with particle numbers of $N=32~768$ (32k), $65~536$ (64k), and $131~072$ (128k) stars.

\begin{figure}[h!]
\begin{center}
 \includegraphics[width=2.5in]{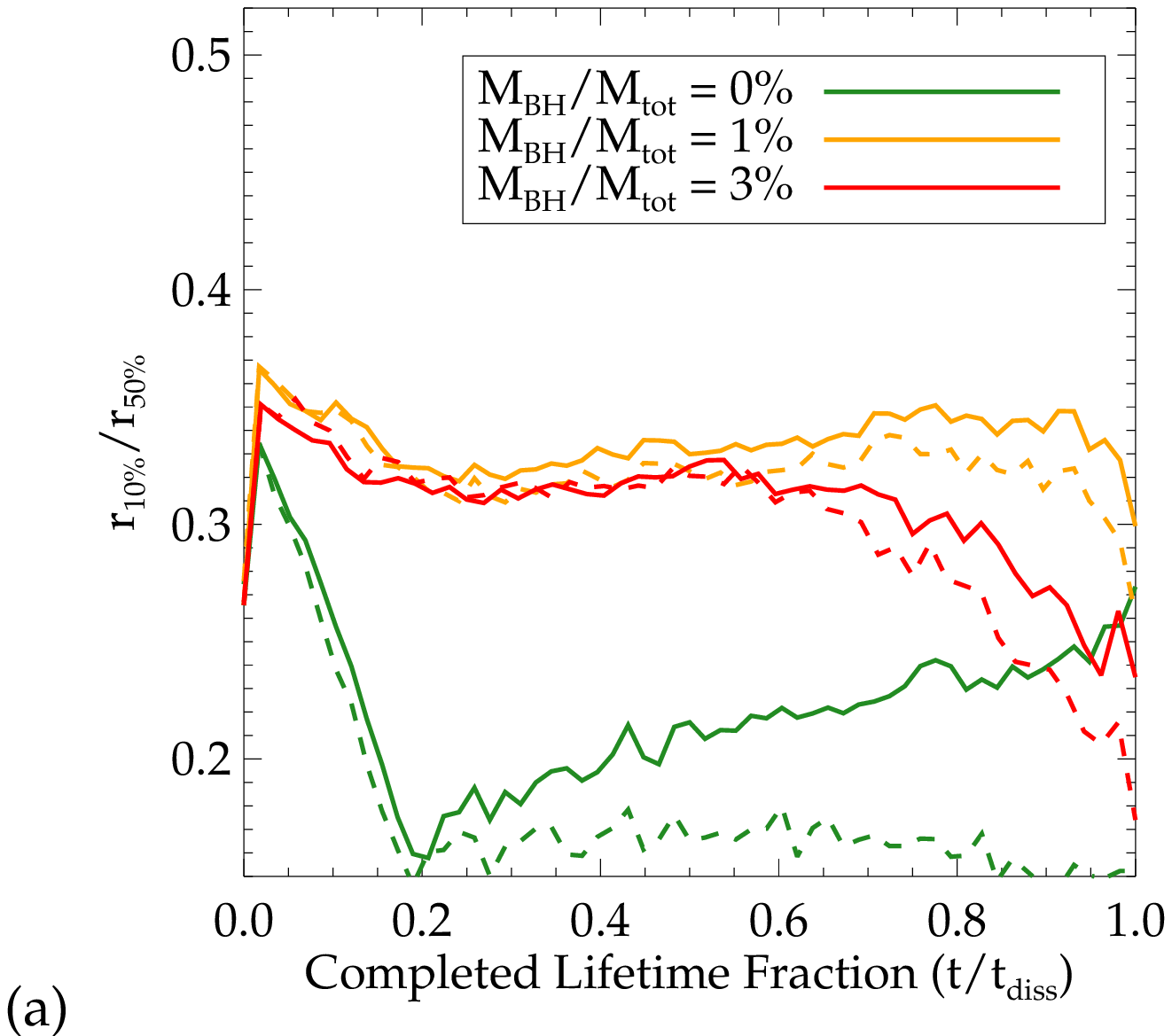} 
  \includegraphics[width=2.5in]{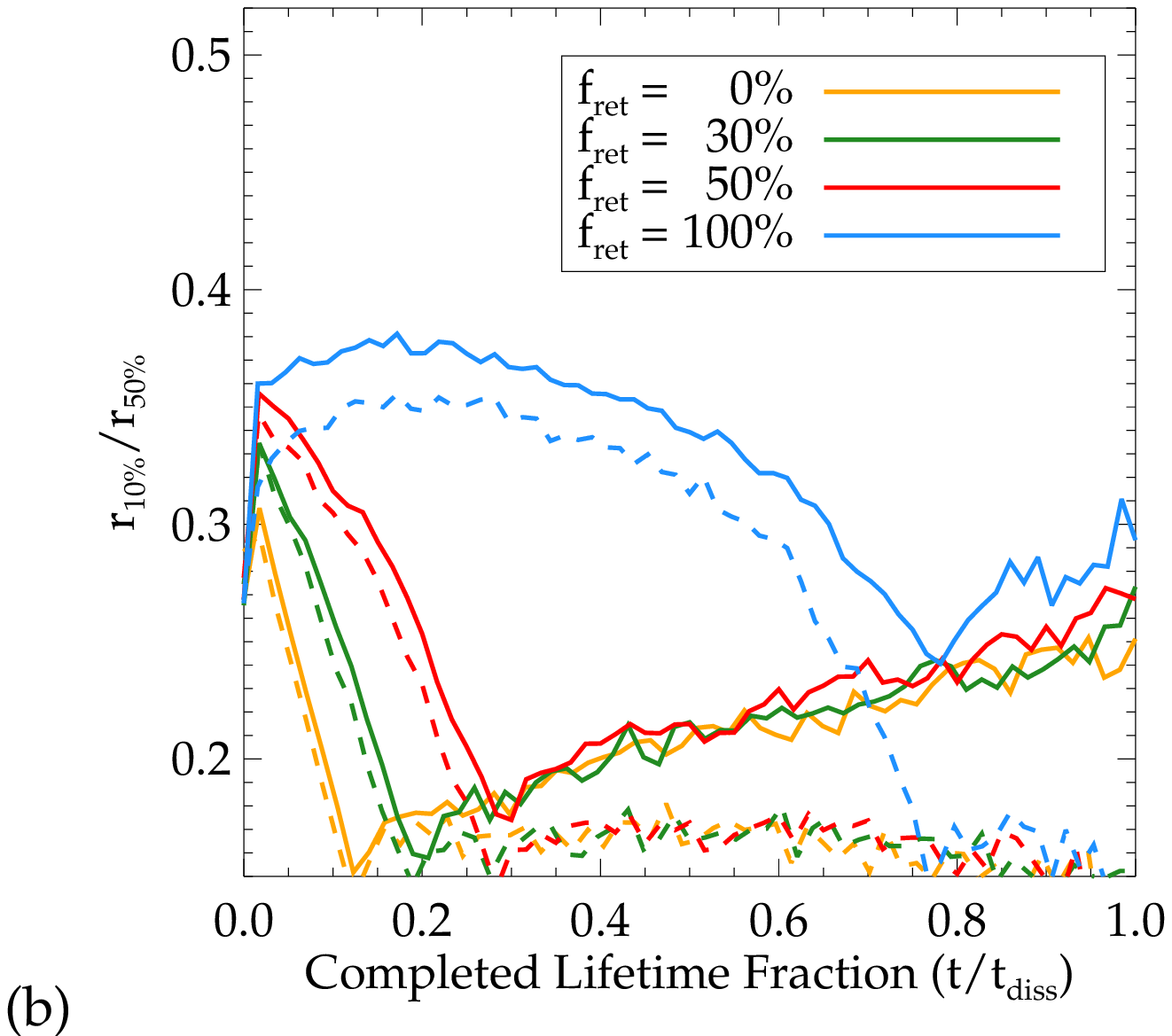} 
 \caption{Characteristic radii ratios ($r_{10\%}/r_{50\%}$, i.e., deprojected radii containing $10\%$ and $50\%$ of the stars) as a function of the completed lifetime fraction for simulations with different a) IMBH masses, b) black-hole retention fractions.  The dashed lines indicate the $r_{L,10\%}/r_{L,50\%}$ (deprojected radii containing $10\%$ and $50\%$ of the mass) evolution when taking the Lagrangian radii.}
   \label{fig:nbody}
\end{center}
\end{figure}

We have investigated the effect of intermediate-mass black holes, stellar-mass black hole retention fractions, and primordial binary fractions on the properties of globular clusters evolving in a tidal field. We studied the effect of the different initial conditions on the cluster lifetime, remnant fraction, mass function, and structural parameters. In addition, we compared the results of the simulations with observational data from the literature and found good agreement. Owing to the specific shape of the King profile, we found the concentration parameter $c$ to be a poor representation of the cluster's internal properties. Especially after core collapse, a King model is not able to reproduce the central cusp in any of our models and the concentration is systematically underestimated. For that reason we also computed the ratio of the radius containing $10\%$ and $50\%$ of the stars in the cluster $r_{10\%}$ and $r_{50\%}$, which is a more accurate quantity than the parametric King fit. Figure \ref{fig:nbody} shows the evolution of the ratio of these two non-parametric radii with time. The core collapse is shown as a prominent dip and is prevented (or delayed) by the presence of an IMBH in the center or a large number of stellar-mass black holes. The simulations showed further that the remnant fraction decreases faster when an IMBH is present in the center. This can be explained with the lower degree of mass segregation which increases the chance of finding high-mass objects in the outskirts of the cluster where they can be easily removed.

\subsection{IFU simulations}

\begin{figure}[h!]
\begin{center}
 \includegraphics[width=5in]{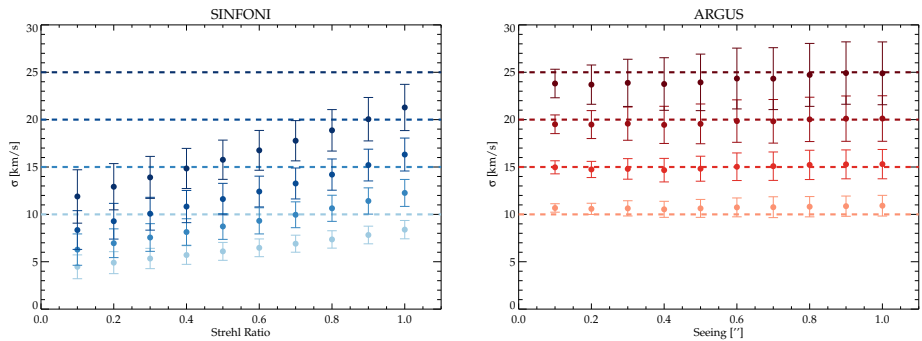}  
 \caption{IFU simulations for SINFONI and ARGUS (central bin) with different input velocity dispersions as a function of Strehl ratio (the amount of light contained in the diffraction-limited core of the PSF, with respect to the total flux) and seeing. The input velocity dispersions are shown as dashed lines, the measured velocity dispersions as dots in the corresponding color.}
   \label{fig:6388}
\end{center}
\end{figure}

Another important part in the field of research on IMBHs is provided by observations of the integrated cluster light. Integral field spectroscopy has become a popular tool in astronomy over the last years. The ability of simultaneously retrieving spatial and spectral data of an object has significant advantages for dynamical and stellar population studies. However, it is not yet clear what is the effect of this observing method on semi-resolved systems such as galactic globular clusters. For this reason it is crucial to simulate the observations in order to understand possible biases. We have developed a tool that allows to simulate IFU observations using star catalogues for different distances and observing conditions (i.e. seeing, Strehl ratio). The program uses a Moffat PSF that is applied to each star and integrated in the grid of the IFU. Using N-body simulations and N-body realizations we have tested the outcome of IFU observations for clusters with different properties, distances, and seeing. The results will be presented in L\"utzgendorf et al. (2015b, in prep). In addition we investigated the specific case of NGC 6388 where two different methods using IFUs brought very different results. While \cite{nora11} found a steeply rising velocity dispersion profile and therefore a strong signature for an IMBH in its center, \cite{lanzoni_2013} measured a central velocity dispersion by taking individual velocities from adaptive optics supported SINFONI observations that is 40\% lower than the value measured with integrated light. We reproduced both observations using the IFU simulation code and find the SINFONI observations being biased towards lower velocity dispersions due to blends of neighboring stars and background light (L\"utzgendorf et al. 2015a, submitted, see Figure \ref{fig:6388}). 

\subsection{AMUSE simulations}

\begin{figure}[h!]
\begin{center}
 \includegraphics[width=5in]{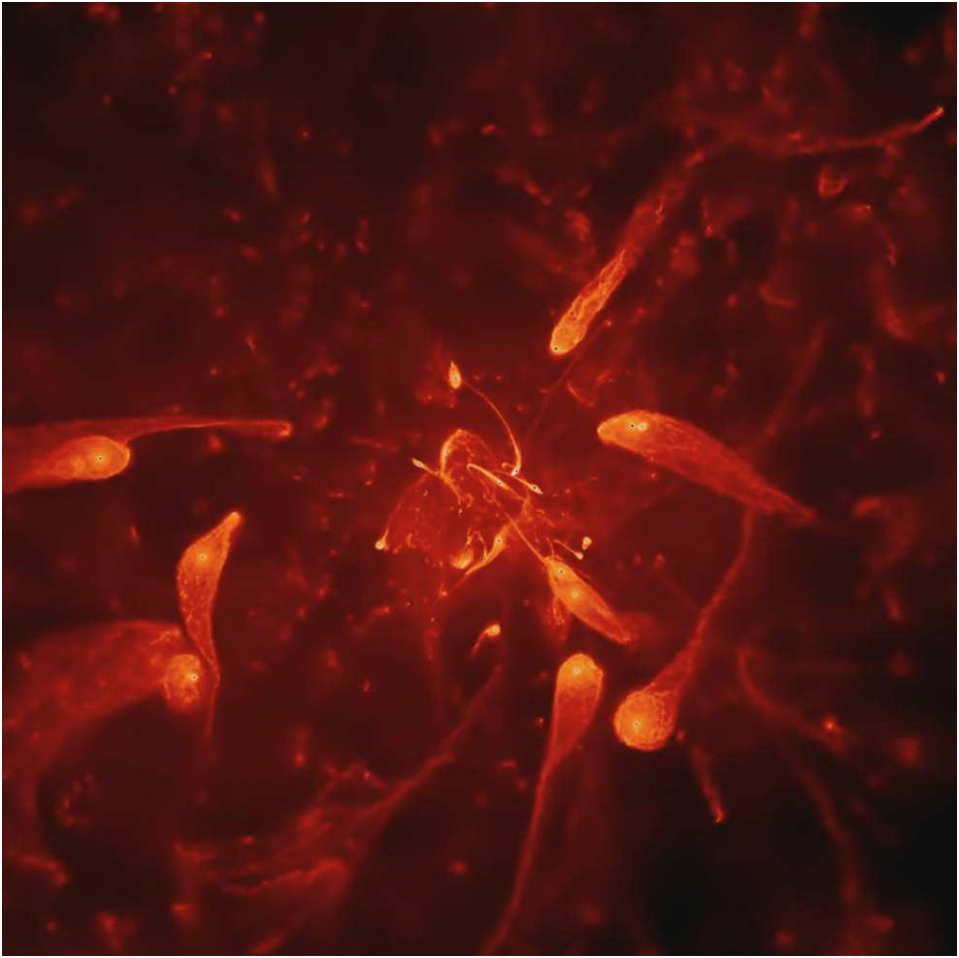}  
 \caption{Snapshot of the AMUSE simulation showing all 27 S-Stars and their stellar winds while orbiting the supermassive black hole in the center (L\"utzgendorf, 2015c, in prep).}
   \label{fig:AMUSE}
\end{center}
\end{figure}

Recently, many IMBH detections were challenged by contradicting measurements from different observing techniques. Especially the discrepancy between kinematic black-hole measurements and the absence of strong X-ray and radio emission from the centers of globular clusters remain an unsolved mystery \citep{strader_2012}. The main uncertainty when translating X-ray and radio flux measurements to black-hole masses is the amount of gas that is accessible for the black hole to accrete. We set out to investigate the effect of stellar winds on the accretion flow of the black hole. By using the Astrophysical Multipurpose Software Environment  \cite[AMUSE,][]{portegies_zwart_2009,portegies_zwart_2013,pelupessy_2013} we combined gravitational physics, stellar evolution and hydrodynamics into a single simulation of stars interacting with a black hole in the center of a globular cluster. The first application of this code will be presented in L\"utzgendorf 2015c (in prep) where the accretion rate of the supermassive black hole in the Milky Way from stellar winds of the surrounding S-Stars is studied (Figure \ref{fig:AMUSE}). The S-Star system in the galactic center is well observed and constrained and provides the ideal laboratory to verify our code. In this work we study the influence of individual stars on the accretion rate as well as temperature and density evolution in the system. The final outcome of the simulations will be compared to accretion rates computed from X-ray observations of the galactic center. From this, we will be able to extend the simulations to actual star cluster sizes and to pin down accretion rates expected from such a system. 

\section{Summary}

We present the work done by our group on the field of IMBHs in globular clusters over the past years. Using integral field spectroscopy in combination of high-resolution HST imaging we have investigated the presence of IMBHs in a sample of 10 galactic globular clusters. We found about a third of them show signatures of an IMBH in the center. Using our data points in combination with literature data we put the IMBH mass estimates in context with properties of their host systems and compared this to existing scaling relations. We found an offset, shallower correlation for the main relations ($\mbh-\sigma$, $\mbh-L$ and $\mbh-M$) which can be explained by the mass loss of globular clusters located in a tidal field over its life time. To support our observations we performed N-body simulations on globular clusters with and without a central IMBH. We found that clusters with IMBHs do not undergo core collapse and lose massive stars such as stellar remnants more rapidly. Further simulations aimed on reproducing IFU simulations by creating fake data sets from realistic N-body simulations and realizations with different distances and observing conditions. The simulations show that internal properties are in general well reproduced with IFU observations but that large uncertainties due to shot noise are unavoidable. In the specific case of NGC 6388 we found a strong bias in the velocity dispersion when computing it from individual velocities due to blend effects and background contamination. Finally, we used the Astrophysical Multipurpose Software Environment (AMUSE) to simulate the accretion of stellar winds onto a supermassive black hole to reproduce the S-Star system in the galactic center. We plan to apply the code developed here to globular clusters and intermediate mass black holes in future work.


\end{document}